**Using deep learning to detect patients at risk for prostate cancer despite benign biopsies**

Bojing Liu[1], Yinxi Wang[1], Philippe Weitz[1], Johan Lindberg[1], Johan Hartman[2], Lars Egevad[2], Henrik Grönberg[1], Martin Eklund[1], Mattias Rantalainen[1,*]

[1]Department of Medical Epidemiology and Biostatistics, Karolinska Institutet, Stockholm, Sweden

[2] Department of Oncology-Pathology, Karolinska Institutet, Stockholm, Sweden

* Corresponding author: mattias.rantalainen@ki.se

**Word counts**

Abstract: 300, Manuscript text: 2542, Figures: 3, Reference: 38

Supplemental eFigure: 2, Supplemental eTable: 3

**Abstract**

**Background:** Transrectal ultrasound guided systematic biopsies of the prostate is a routine procedure to establish a prostate cancer (PCa) diagnosis. However, the 10-12 prostate core biopsies only sample a small volume of the prostate, and tumour lesions in regions between biopsy cores can be missed, leading to low sensitivity to detect clinically relevant cancer. We hypothesise that computer models can detect patients with PCa despite benign biopsies. As a proof-of-principle,we developed and validated a deep convolutional neural network (CNN) model to distinguish between morphological patterns in benign prostate biopsy whole slide images from men with and without established cancer.

**Methods:** This study included 14,354 hematoxylin and eosin stained whole slide images from benign prostate biopsies from 1,508 men in two groups: men without established PCa diagnosis and men with at least one core biopsy diagnosed with PCa. 80% of the participants were assigned as training data and used for model optimization (1,211 men), and the remaining 20% (297 men) as a held-out test set to evaluate model performance. An ensemble of 10 deep CNN models was optimized for classification of biopsies from men with and without established cancer.

**Results:** Area under the receiver operating characteristic curve (ROC-AUC) was estimated as 0.727 (bootstrap 95% CI:0.708-0.745) on biopsy level and 0.738 (bootstrap 95% CI:0.682-0.796) on man level. At a specificity of 0.90 the model had an estimated sensitivity of 0.348.

**Conclusion:** The developed model can detect men with risk of missed PCa due to under-sampling of the prostate. The proposed model has the potential to reduce the number of false-negative cases in routine systematic prostate biopsies and to indicate men who could benefit from magnetic-resonance guided re-biopsy.

**Introduction**

Prostate cancer (PCa) is the second most common cancer in men worldwide[1]. Transrectal ultrasound (TRUS) guided 10-12 core needle prostate biopsy is the routine diagnostic tool for men who have elevated prostate-specific antigen (PSA) and/or abnormal digital rectal examination (DRE) and/or other suspicious indications[2]. A major limitation of the systematic TRUS biopsy is undersampling, causing cancer lesions between biopsy cores to be missed[3]. Consequently, the sensitivity of TRUS biopsy has been reported as low as 32-58%[4,5] and increasing the number of biopsy cores only marginally improves the sensitivity and mainly detects indolent cancers[6]. Although magnetic resonance (MRI)-guided targeted biopsy is recommended for biopsy naive patients and patients with indications for repeated biopsy, it is not widely available to patients on a broad scale. Improvements in the sensitivity of TRUS biopsy, especially for more aggressive PCa, is therefore needed to improve the detection of clinically relevant cancer and to reduce unnecessary rebiopsies.

PCa diagnosis based on histopathological inspection of prostate biopsy (hematoxylin-eosin (H&E) stained) slides is challenging and prone to inter-assessor variability [3,7]. Currently, there are three features (perineural invasion, glomerulations, and mucinous fibroplasia) not observed in benign prostate glands and therefore are considered to be diagnostic for PCa. However, most PCa are identified based on a combination of non-specific major and minor cancer architectural and cytological features[3]. The major features (eg. infiltrative growth pattern, absence of basal cells, and nuclear atypia) are strongly linked to cancer, while the minor features (eg. cytoplasmic amphophilia, intraluminal contents, mitosis and apoptosis) have a weaker link to cancer, and can also be seen in non-cancer lesions[3]. In addition, premalignant changes of high-grade prostatic intraepithelial neoplasia (HGPIN) and atypical small acinar proliferation (ASAP) are associated with later PCa diagnosis[3]. It is well recognized that development of cancer is a continual process in which cells gradually become malignant as a result of the accumulation of mutations and selection[8]. Moreover, epigenetic changes have been reported in the early development of cancer and distinct epigenetic signatures may be present in neighbouring prostatic tissues adjacent to the PCa foci[9]. Therefore, subtle cancer-related morphological structures with clinical relevance, which are either non-specific or not detectable by human eyes, could be present in benign cores sampled in the vicinity of the cancer areas.

Deep learning (DL)[10] in the form of convolutional neural networks currently offers state-of-the-art performance for image classification. DL has recently demonstrated human level performance in routine pathology tasks[11], including cancer detection and grading[12,13]. DL has also demonstrated ability to predict additional factors from H&E stained WSIs, which cannot be determined in routine pathology by a human assessor, including status of molecular markers[14,15] and microsatellite instability[16]. Consequently, we hypothesize that there might be subtle morphological patterns present in prostate H&E WSIs that can be modelled by DL to predict clinically relevant factors. In this study, we investigate the potential to apply DL to recognize cancer-associated morphological changes in benign prostate biopsies, and assess to what extent such models could improve sensitivity to detect clinically relevant PCa in TRUS guided systematic prostate biopsies.

## Material and Methods

### Study population

The study was based on a subset of men from the STHLM3 study[17], a prospective population-based diagnostic study for PCa. The study enrolled a random sample of men aged between 50 and 69 years who were free of promoted PCa from the Swedish population register. Out of those, 59,159 men accepted to participate in the study[17] and 7,406 underwent TRUS prostate biopsies (10-12 cores). The study protocol was reviewed and approved by local and governmental ethics committees. All participants provided written informed consent. The biopsies were reviewed and graded according to the International Society of Urological Pathology (ISUP)[18] system by a single pathologist (Lars Egevad). The study included 606 men with detected cancer (ISUP≥1) randomly sampled together with 134 men with only benign biopsies, for details see[12], all benign biopsies from these 740 men were included in the present study. To further increase the sample size, all benign biopsies from participants enrolled during 2015 in STHLM3 (N=877 men, 585 benign, 292 cancer) were also scanned and included.

To reduce the risk for labeling error, we excluded participants or biopsies if there were suspected errors in gleason score (GS) (i.e. duplicates with inconsistent patient level GS), cancer diagnosis, cancer length. Patients with missing in PSA and age were excluded in the analysis (**eFigure 1**). Only benign biopsy cores were included. In total, 14,354 benign biopsies from 829 men (6523 biopsies) with detected cancer (at least one core biopsy had cancer) (cancer-benign biopsies), and 679 men (7831 biopsies) with no detected cancer (all cores classified as benign) (benign-benign biopsies) were included. All WSIs were scanned at 20X magnification (Hamamatsu Nanozoomer-XR, pixel size=0.4536μm). The dataset was randomly split on patient level into training (973 men), validation (238 men), and held-out test (297 men) sets. The split (i.e. training and validation versus held-out test) was balanced on age, PSA levels, and cancer diagnosis. For cancer patients, the split was further balanced on ISUP grade and the length of the cancer in the biopsy (**eTable 1** Sample description). The study was approved by the Stockholm regional ethics board.

**Data pre-processing**

**Tissue segmentation and image tiling**

WSIs were read using OpenSlide[19]. Tissue areas in WSIs were segmented against background, downsampled by a factor of 2 (10X magnification level, pixel size approximate 0.90 um) and split into tiles of size 299x299 pixels, with 50% overlap. In total, we obtained 8,780,026 tiles for training, 2,192,124 for validation, and 2,713,314 for the test set.

**Image classification**

We applied a deep convolutional neural network model based on the ResNet18 architecture[20,21] for binary classification of benign biopsy cores from benign men versus benign biopsy cores from men with detected cancer, using the preprocessed image tiles together with age and PSA as predictors (**Figure 1**). Age was categorised in five groups <= 55, 55- 60, 60-65, 65-70, 70-100 years. PSA level at the biopsy was grouped as 1-3, 3-5, 5-10, >=10 ng/ml. The model was optimized using the training set as a weakly supervised binary classifier, with a patient-level label assigned to each tile. Model weights were initialized from weights pre-trained on ImageNet[22]. We applied binary cross-entropy loss and the Adam optimizer[23]. Hyper-parameters were optimized using cross-validation in the training set, including tile size, magnification level, and learning rate. An ensemble of 10 CNN base-models was used to further reduce variance in predictions. Classification performance was evaluated in the validation set and the held-out test set using Receiver Operating Characteristics (ROC) and Area Under the ROC curve (AUC). Sensitivity for detecting cancer on patient level was evaluated at specificity levels of 0.99, 0.98, 0.95, and 0.90.

**Model Interpretation**

We randomly selected 20% benign tiles stratified by PCa diagnosis from the test set and applied Uniform Manifold Approximation and Projection (UMAP)[24] to generate a two-dimensional representation of the 512-dimensional feature vector in the DL model. Based on the UMAP representation, we selected four distinct regions in each class for visualisation of tiles. For each region, we sampled 16 cancer-benign tiles with highest predicted score and 16 benign-benign tiles with the lowest predicted score to illustrate morphological structures captured by the model. Additional visualisation of key patterns in the sample tiles were performed by guided gradient-weighted class activation mapping (Guided Grad-CAM)[25,26]. See Supplementary Methods for further details.

**Data Availability**

The data analysed are part of the STHLM3 study, data cannot be distributed without access control due to local privacy laws. Requests for data access can be submitted to the STHLM3 study for consideration after ethical permission has been obtained.

**Code availability**

All data analyses are based upon publicly available software packages (see Methods).

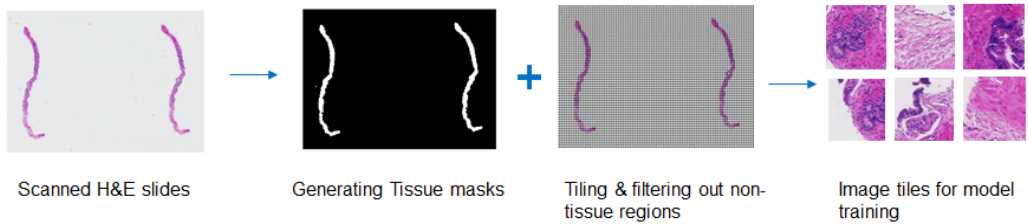
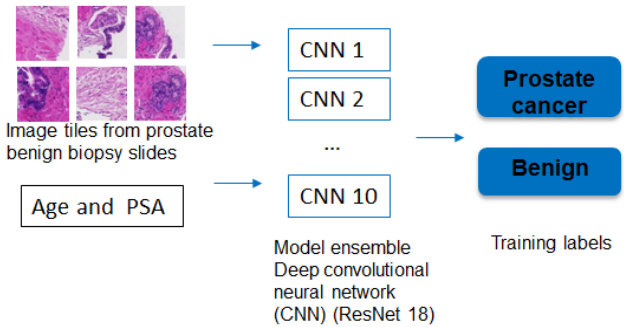
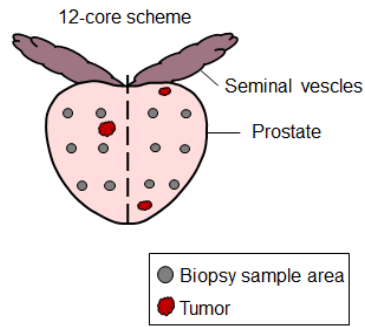

**Figure 1. Overview of image pre-processing and DL modelling for detecting cancer patients from benign prostate biopsies only**. a) Pre-processing of WSIs. b) Schematic overview of the image classification by a deep CNN ensemble. c) Systematic prostate biopsy 12-core scheme.

## Results

The CNN ensemble model was optimised on 9,192 prostate biopsy WSIs (8,780,026 tiles) from 973 men (535 cancer, 438 benign) in the training set. Classification performance was first estimated in the validation set (130 PCa and 108 benign men, 2,311 WSIs, 2,192,124 tiles) and subsequently evaluated in the held out test set (164 PCa and 133 benign men, 2851 WSIs, 2,713,314 tiles) (**Figure 2a** - ROC curve). We observed a tile level prediction performance of AUC=0.701 (95% $CI_{bootstrap}$:0.700-0.701), slide level AUC=0.727 (95% $CI_{bootstrap}$:0.708-0.745), and a patient level AUC=0.738 (95% $CI_{bootstrap}$:0.682-0.796) in the held-out test set. Based on patient level analysis, we assessed prediction performance at specificity 0.99, 0.98, 0.95, and 0.90. (**Figure 2b**). For overall cancer detection, we observed a sensitivity of 0.043 (95% $CI_{bootstrap}$: 0.000 - 0.213) at the specificity of 0.99, a sensitivity of 0.177 (95% $CI_{bootstrap}$: 0.038 - 0.240) at the specificity of 0.98, and a sensitivity of 0.224 (95% $CI_{bootstrap}$: 0.153 - 0.311) at specificity of 0.95, and a sensitivity of 0.348 (95% $CI_{bootstrap}$: 0.205 - 0.468) at the specificity of 0.90. These sensitivities differed by ISUP with higher sensitivity for more aggressive cancers graded as ISUP 4 or 5 (**Figure 2b** and **eTable 2**). We applied logistic regression to establish baseline classification performance using only age and PSA as predictors and observed an AUC=0.562 (95% CI$_{bootstrap}$:0.510-0.612) in the held-out test set.

To characterize the CNN model further, we applied UMAP to generate a 2-D projection of the 512 high dimensional feature representation in the CNN. Benign tiles from men without PCa and benign tiles from men with PCa had clear differences in tile density distribution in the UMAP projections (**Figure 3a & b** middle plots). Image tiles from selected regions in UMAP projections displayed differences in morphological features. We also applied the guided Grad-Cam[25,26] method to further visualize areas in tiles linked to cancer associated morphological patterns. As expected, benign tiles from men with benign diagnosis (**Figure 3a**) generally had fewer areas and weaker indication, while tiles from PCa men had more and larger areas, and with stronger indication (**Figure 3b**) highlighted.

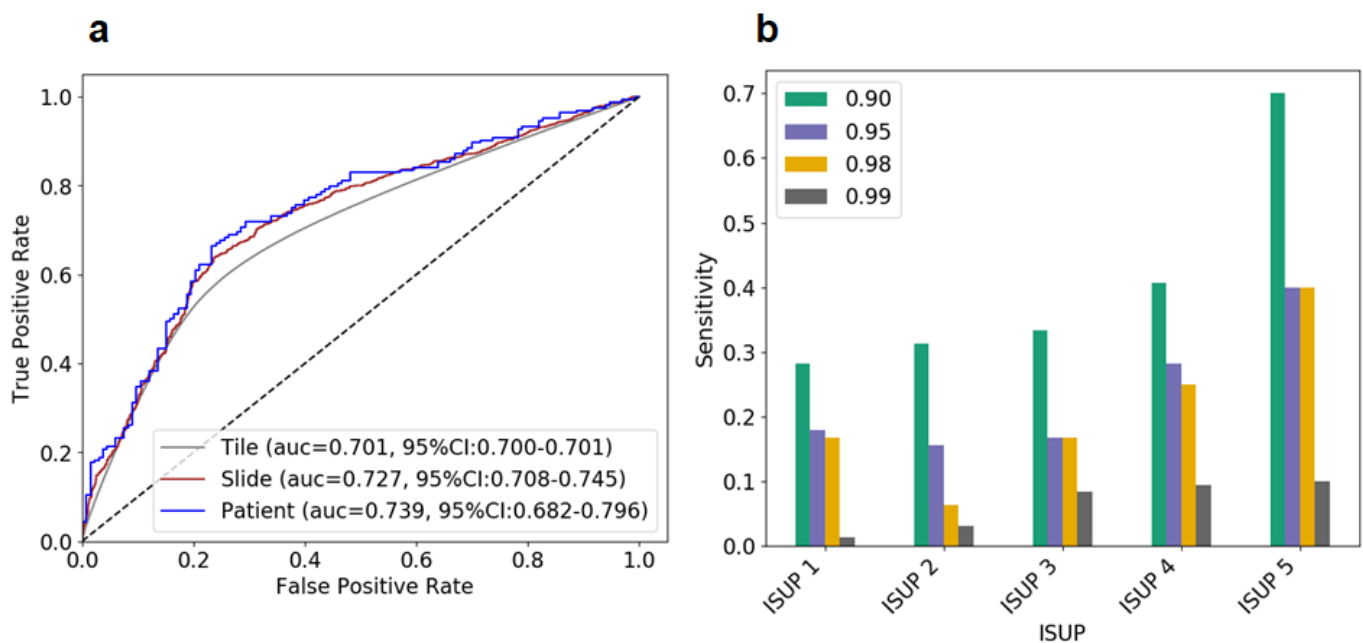

**Figure 2. Prediction performance in held-out test set** a) Receiver Operating Characteristic (ROC) curves and associated AUC estimates in the hold-out test set for prediction of cancer versus benign PCa diagnosis on tile, slide, and patient level; b) corresponding sensitivity by ISUP group at specificities of 0.99, 0.98, 0.95, and 0.90.

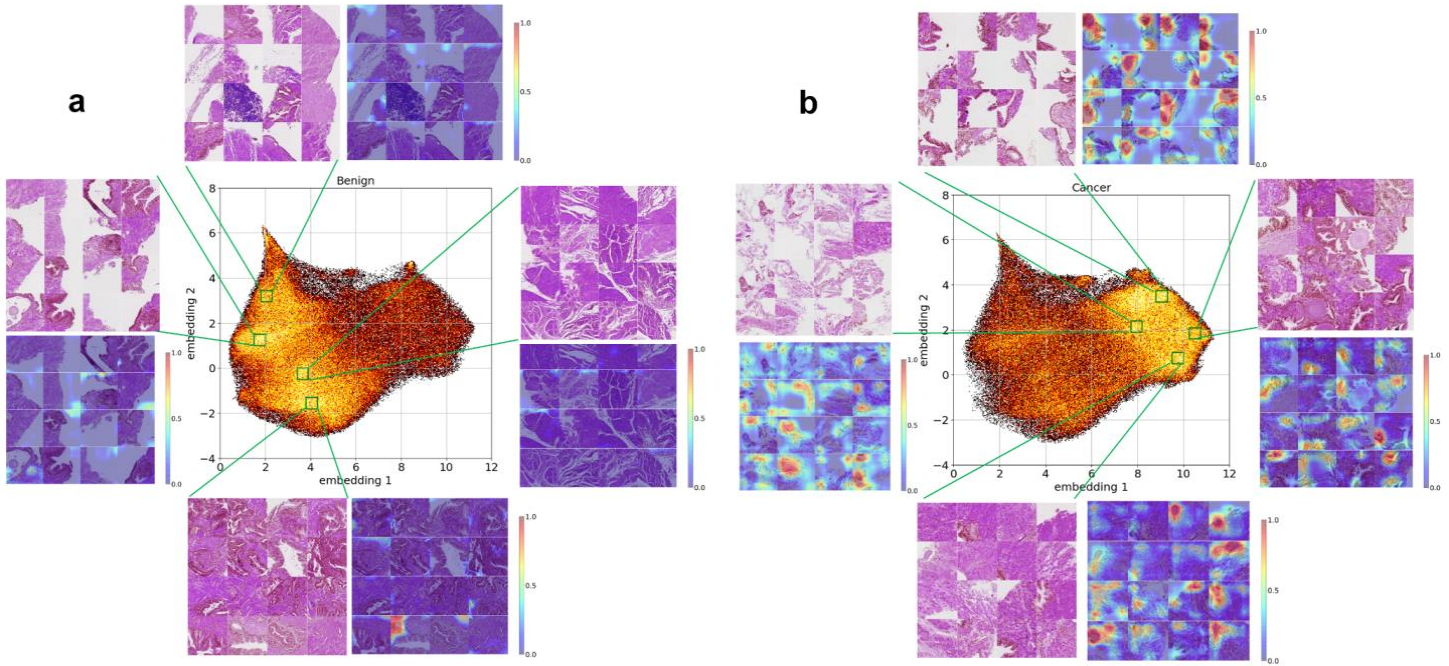

**Figure 3. Model characterization and interpretation using UMAP and grad-cam** a) UMAP projection of benign tiles from men with benign diagnosis (middle). Examples of benign tiles and corresponding grad-cam (outside) from four distinct regions from the UMAP. b) UMAP projection of benign tiles from men with PCa diagnosis (middle). Examples of benign tiles and corresponding grad-cams (outside) from four distinct regions from the UMAP. The two-dimensional plots are reduced from 512-dimensional feature maps from the last fully connected layer.

**Discussion**

In this study we developed a DL-based model for classification of WSIs of benign prostate core biopsies from men without a cancer diagnosis and from men with at least one biopsy core with cancer. The model achieved a patient level AUC of 0.738, and given a specificity of 0.90, the proposed model was able to detect 34.8% of all PCa, and 41% ISUP 4 and 70% of ISUP 5 PCa, from benign biopsy cores only.

There is a plethora of challenges associated with PCa detection by biopsy. PCa is often multifocal and systematic TRUS biopsies only samples a small volume of the prostate, hence, there is an intrinsic and substantial risk that biopsy cores miss a cancer lesion[27]. The peripheral zone that harbors 70-80% of the PCa are most frequently sampled, while the transition zone containing 20% of the cancers is typically not sampled during the initial biopsy[3]. Sampling templates in clinical practice also vary substantially. A recent study reported that 1/3 of 137 urologists never or seldomly sampled the anterior part of the prostate[28] despite its high cancer frequency[29]. The possible false-negatives from the initial biopsy require repeated biopsies for patients with abnormal DRE or continuously increased PSA or PSA derivatives. However, due to the nonspecificity of these indicators, cancer detection rate for the second biopsy under the saturation protocol (>= 20 cores) remains low (22-24%)[30,31]. In fact, multiple re-biopsies may be performed before establishing a PCa diagnosis[6]. The procedure causes patients' anxiety and potential side-effects, including sepsis[6].

Our DL model identified 34.8% of PCa cases at a specificity of 0.90 from only benign biopsies, including 70% of ISUP 5 and 41% of ISUP 4 cases. This suggests that there exist distinct, albeit subtle, morphological differences between benign prostate tissue in men with cancer present in the prostate (adjacent, or nearby) and men without any cancer present. Hence, subtle differences can be captured by image analysis using DL, which are typically not captured in routine histopathology assessments by human assessors. A high predicted risk score from a computer model could thus potentially assist pathologists, and indicate either further review of existing biopsies, or guide decision for re-biopsy. Notably the model showed higher sensitivity for more aggressive PCa (ISUP 4 and 5) compared to low-to-medium risk PCa, although undersampling of cancerous tissue may be less prevalent in poorly differentiated PCa since larger volume tumours are more often seen in more advanced GS patterns[32]. PSA-based screening and subsequent biopsy reduce the prostate-specific mortality, however, it leads to overtreatment of low risk ISUP 1 cancer[33]. In fact, 50% of men diagnosed by systematic biopsy have indolent low-risk PCa[34], which may only require active surveillance rather than immediate intervention[35]. The model has desirable properties to contribute to the identification of high risk PCa cases missed due to sparse sampling of the prostate during biopsy, while it detects low risk PCa (low ISUP) at lower rate, thus not driving the risk of additional over-treatment. Despite previous success of DL in detection of cancer, cancer grading, and determination of cancer length and prognostic indicators[12,14,36–38], there have been few attempts to extract novel information beyond what human assessors can detect in WSIs. We provide the first evidence that deep CNN models can distinguish between benign tissue from men with established cancer, and benign tissue from men with only benign biopsies.

The study has several strengths. The study is based on a large sample of 14,354 WSIs (from 1,508 men) digitized on a single scanner model, precluding potential bias due to use of multiple scanning devices. The sample was part of the well-controlled STHLM3 population-based study. The biopsy procedure was standardized and the pathology report for PCa diagnosis was centralized and blinded for both urologists and the pathologist regarding clinical characteristics. The study has some limitations, benign biopsies with cancer were taken from men that were not actually missed, since they had at least one positive biopsy. Ideally, we would investigate biopsies from men with initial negative biopsies that developed cancer later on. While the well-controlled study design and scanning protocol provides high quality data and opportunity to demonstrate the feasibility of the approach on a conceptual level, it does not demonstrate generalisability to other populations. Future studies would have to be focused on determining to what extent the model generalizes to other settings. Furthermore, the study was enriched with higher ISUP PCa, which could contribute to the observed higher sensitivity to detect men with high ISUP PCa.

**Conclusions**

Subtle tumour-associated morphological changes in benign prostate tissue that cannot be distinguished by human eye can be captured by deep CNN models trained on large numbers of benign prostate biopsy histopathology images. The developed model has the ability to detect men with risk of missed prostate cancer after biopsy, caused by under-sampling of the prostate. The proposed model has the potential to reduce the number of false negatives caused by sparse sampling of the prostate volume in routine systematic prostate biopsies and to indicate men that could benefit from MRI-guided re-biopsy.

**Acknowledgment**

This project was supported by funding from the Swedish Research Council (MR, BL), Swedish Cancer Society (Cancerfonden) (MR, ME), Swedish e-science Research Centre (SeRC) - eCPC (MR).

**Competing interests**

The authors declare that there are no competing interests.

**Author contribution**

conception and design (MR), acquisition of data (MR, JH, HG, LE), data analysis and interpretation of data (BL, YW, PW, MR), drafting of the manuscript (BL), statistical analysis (BL), obtaining funding (MR), administrative, technical, or material support (MR), supervision (MR), critical revision of the manuscript for important intellectual content (all authors).

# Supplementary materials

**Image pre-processing**

*Tissue segmentation*

We first generated tissue masks separating the prostate tissue from the background. The images were read at full resolution (0.4536 um /pixel), downsampled by a factor of five from the 20X images using the resolution pyramid from OpenSlide (v. 1.1.1)[1]. Next, we converted the RGB images to HSV encoded images using the color module in scikit-image (v.0.14.2)[2]. Tissue areas were defined as regions above the Ostu's threshold in saturation and above 0.75 threshold in hue, background areas were discarded from further analysis. In addition, pen marks were present in a small number of benign slides, which marked suspicious regions for cancer or prostatic intraepithelial neoplasia (PIN). The pen marks were filtered out from the tissue areas using the following functions from OpenCV (v.3.4.2)[3] : first, the RGB channel was converted to grayscale using COLOR_BGR2GRAY(); second, a "2D" Laplacian operator was applied using Laplacian () with the Sobel kernel size of three and depth of the output image size set as CV_16S; last, the absolute value of the resulting response was obtained using convertScaleAbs () and the pen marks subsequently filtered if below the threshold 20 of the absolute value. We further smoothed tissue masks (removing peppers and salts) by applying a morphological opening and losing, with a disk with a radius of six using functions of disk (), opening(), and closing() in scikit-image morphology module[4].

*Image tiling and quality control*

To facilitate model training, WSIs were tiled into small image patches. We used a window size of 598 × 598 pixels from the 20X full resolution (0.4536 um /pixel) and a stride of 299 pixels. The tiles were subsequently downsampled by a factor of 2, the resulting image patches are of size 299 × 299 pixels and with 50% overlap between the adjacent tiles. Tiles that didn't contain more than 20% of tissue were excluded. To ensure the quality of the images, we excluded unsharp tiles due to poor focusing during scanning. We calculated the variance after the Laplacian filter using variance_to_laplacian () function in OpenCV[3] and excluded tiles with the variance lower than 200.

**Model development**

*Model overview*

We randomly split the dataset on the man level into training (973 men), validation (238 men), and held-out test

(297 men) sets. Training and validation sets combined, and the held-out test, were balanced on PCa diagnosis, age, PSA, and further balanced on ISUP for PCa patients. The validation set was used only once to validate final model performance, and the held-out test was only used once to evaluate final model performance. We used the ResNet-18[5,6] with weights initialized from a model pre-trained by ImageNet [7] images as the base model. This output layer was subsequently concatenated with categorical covariates of age (<= 55, 55- 60, 60-65, 65-70, 70-100 years) and PSA (1-3, 3-5, 5-10, >=10 ng/ml). We additionally included a fully connected layer to 256 hidden units allowing potential interactions among age, PSA and the learned image features. We applied 50% drop-out after global-average-pooling of the final convolutional layer as well as 50% drop-out and max-norm regularization[8] in the last fully connected layer of the model to mitigate potential over-fitting.

*Model training*

Parameters from all layers were optimized by backpropagation through mini-batch stochastic gradient descent (SGD) and binary cross-entropy loss. Learning rates of 1e-5, 1e-6, and 1e-7, tile size of 299 by 299 pixels or 598 by 598 pixels , and magnification levels of 20X and 10X were evaluated using five-fold cross validation in the training set. Tile size of 299 by 299 pixels at the magnification level of 10X and learning rate of 1e-5 were selected for subsequent optimizations and analyses. The initial learning rate of 1e-5 was reduced by 50% if loss function stopped improving (measured by minimum change of 0.0001) for consecutive 35 epochs. We applied the standard data augmentation including rotation and flip. In order to reduce overfitting, for each fold, the training set was split into training, tuning, and inner validation sets, which ensured that the best model was selected based on the performance in the tuning set and model performance was validated in the inner validation set.

To improve the model generalization, we used an ensemble of ten CNN models sharing the same model architecture and hyperparameters, while the variability within the ensemble arises from the stochastic sampling during training. For training, at each step we randomly sampled a mini-batch of 170 images with equal proportion of image tiles from PCa and non-PCa groups. During optimization we defined 300 iterations as one 'partial epoch', after which the tuning set was evaluated. We allowed a maximum of 400 partial epochs and used an early-stopping with patience of 35 epochs. Less than a minimum change in loss of 0.0001 was considered as no improvement.

Computations were performed on a server with four Nvidia RTX 2080 Ti graphics processing unit (GPU) cards. Images were imported to the Python environment using OpenSlide (v. 1.1.1). Convolutional neural networks were implemented in Python 3.6.4 using Keras (v. 2.2.4) with TensorFlow (v. 1.11.0) backend.

**Model prediction**

The final model prediction performance was only validated once in the validation set (**eFigure 2** and **eTable 3**) and only tested once in the held-out test set using Receiver Operating Characteristics (ROC) and Area Under the ROC curve (AUC), and 95% confidence intervals for AUC were obtained using bootstrap (2,000 bootstrap samples). Tile level predictions in the validation and test sets were obtained by averaging of the prediction scores from the ten-model ensemble. The slide level predictions were defined by the were aggregated using the 75th percentile of the predicted class probabilities across all tiles from each WSI. Patient level predictions were defined based on the median across all WSI level predictions of a patient individual. Percentiles for tile-to-slide and slide-to-patient aggregations were optimized using five-fold cross validation.

**Model interpretation**

*Uniform Manifold Approximation and Projection*

For model interpretation, we selected the model from the ten ensemble models with the highest tile level AUC obtained in the testset. We randomly selected 20% tiles in the test set stratified by PCa diagnosis and generated 2-D UMAP from the modelled 512-dimensional feature vector using the "umap-learn" [9] with parameters set to n_neighbors=15, min_dist=0.1, n_components=2, metric='euclidean'. Through visualization of the UMAP, we selected four distinct regions for PCa patients and healthy men representing characteristic morphological structures learned from CNN.

*Guided gradient-weighted class activation mapping (Guided Grad-CAM)*

We applied guided gradient-weighted class activation mapping (Guided Grad-CAM)[8,10] to visualize salient morphological features underlying the DL-based prediction. For each above selected region from the UMAP, we generated grad-cams visualizations for 16 tiles with the highest prediction score among PCa patients and 16 tiles with the lowest prediction among healthy men to visualize the morphological patterns linked to PCa prediction. We calculated the gradients of cancer with respect to the 8 by 8 feature map activations from the last convolutional layer (ie. 8 × 8 × 512 dimension). The gradients for each 8 by 8 feature map were global-average-pooled to obtain feature-map-specific weights. The weighted linear combination of the 512 forward activation maps was followed by a ReLU function to obtain a 8 by 8 coarse heatmap. This heatmap was further upsampled and interpolated to the corresponding image tile size (299 by 299 pixels) to visualize morphological features that were potentially linked to prostate cancer.

**eTable 1.** Sample distribution

|  | **Training set (training + validation)** | **Held-out test set** | **p-value** |
|---|---|---|---|
| **Prostate cancer, N(%)** |  |  |  |
| No | 546 (0.55) | 133 (0.58) |  |
| Yes | 665 (0.45) | 164 (0.45) | 0.976[1] |
| ISUP 1 | 325 (0.27) | 78 (0.26) |  |
| ISUP 2 | 121 (0.10) | 32 (0.11) |  |
| ISUP 3 | 51 (0.04) | 12 (0.04) |  |
| ISUP 4 | 127 (0.10) | 32 (0.11) |  |
| ISUP 5 | 41 (0.03) | 10 (0.03) | 0.999[1] |
| Mean cancer length, mean(±SD) | 0.43 (±0.82) | 0.41 (±0.85) | 0.434[2] |
| Maximum cancer length, mean(±SD) | 2.67 (±3.77) | 2.58 (±3.66) | 0.462[2] |
| **Age (mean±SD)** | 63.52 (±5.30) | 63.14 (± 5.37) | 0.119[2] |
| **PSA(ng/ml), N(%)** |  |  |  |
| [1,3) | 361 (0.30) | 91 (0.31) |  |
| [3,5) | 508 (0.42) | 125 (0.42) |  |
| [5,10) | 256 (0.21) | 63 (0.21) |  |
| [10,1e+04) | 86 (0.07) | 18 (0.06) | 0.934[1] |

1.Chisq test, 2. Mann-Whitney U test

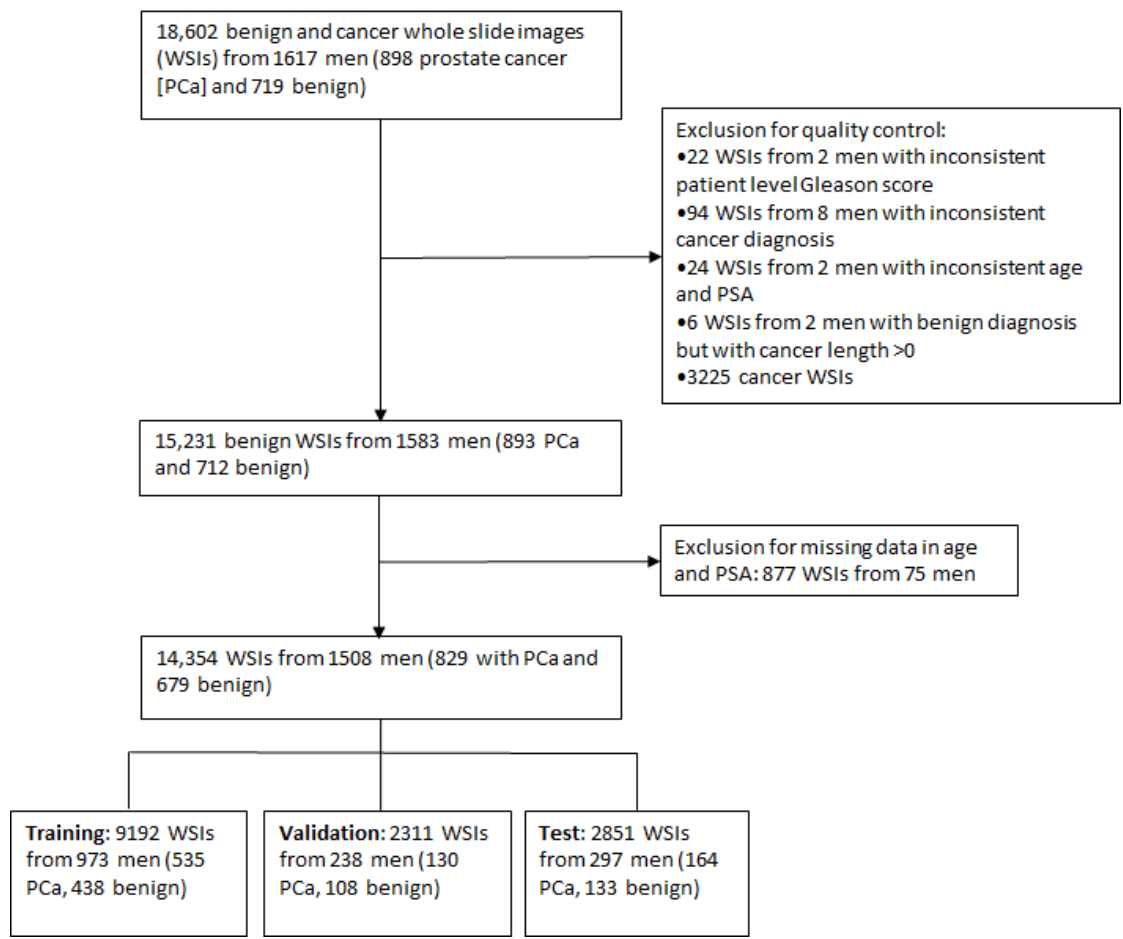

**eFigure 1.** Flow-chart

**eTable 2** Sensitivity by ISUP according to different specificity levels in held-out test set

|             | Sensitivity |        |        |        |        |
| ----------- | ------ | ------ | ------ | ------ | ------ |
| **Specificity** | ISUP 1 | ISUP 2 | ISUP 3 | ISUP 4 | ISUP 5 |
| 0.99        | 0.01   | 0.03   | 0.08   | 0.09   | 0.10   |
| 0.98        | 0.17   | 0.06   | 0.17   | 0.25   | 0.40   |
| 0.95        | 0.19   | 0.16   | 0.17   | 0.31   | 0.40   |
| 0.90        | 0.28   | 0.31   | 0.33   | 0.41   | 0.70   |

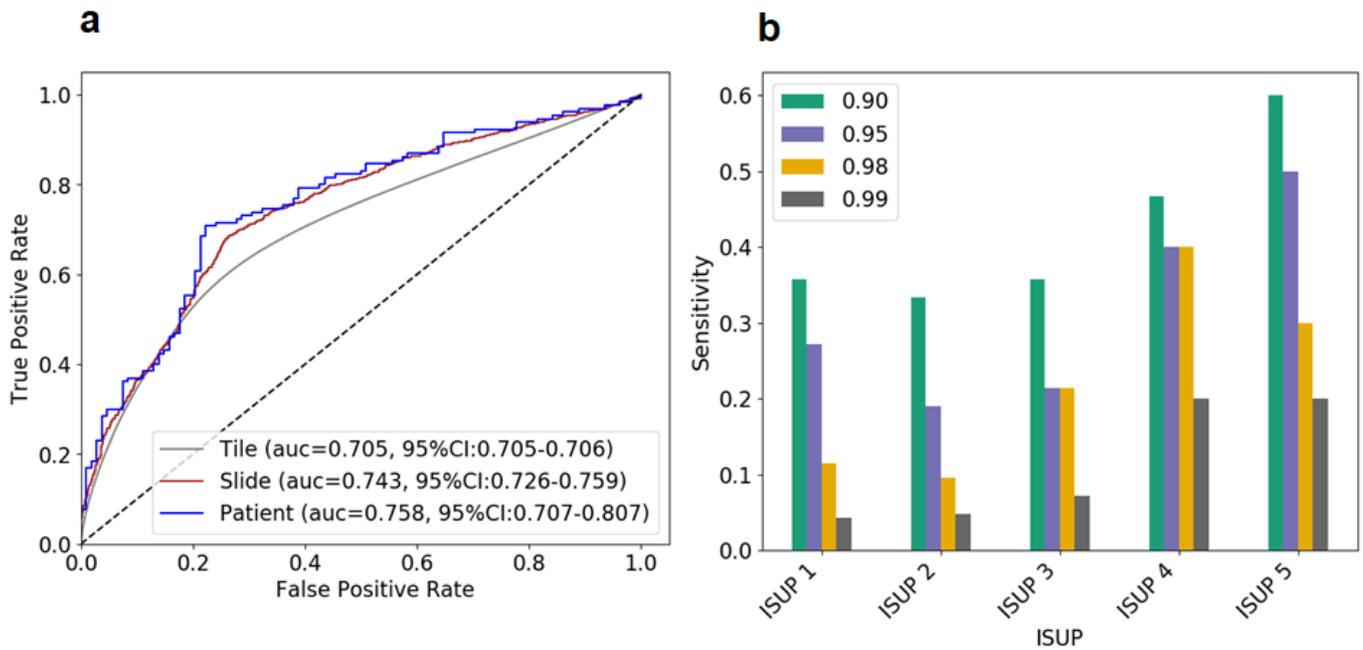

**eFigure 2. Prediction performance in validation set** a) Receiver Operating Characteristic (ROC) curves and associated AUC estimates in the validation set for prediction of cancer versus benign PCa diagnosis on tile, slide, and patient level; b) corresponding sensitivity by ISUP group at specificities of 0.99, 0.98, 0.95, and 0.90.

eTable 3 Sensitivity by ISUP according to different specificity levels in validation set

|  | Sensitivity | | | | |
|---|---|---|---|---|---|
| **Specificity** | ISUP 1 | ISUP 2 | ISUP 3 | ISUP 4 | ISUP 5 |
| 0.99 | 0.04 | 0.05 | 0.07 | 0.20 | 0.20 |
| 0.98 | 0.11 | 0.10 | 0.21 | 0.40 | 0.30 |
| 0.95 | 0.27 | 0.19 | 0.21 | 0.40 | 0.50 |
| 0.90 | 0.36 | 0.33 | 0.36 | 0.47 | 0.60 |